# Gradient-free temperature control over micrometric areas for thermoplasmonic applications in micro-and nano-devices.

Javier González-Colsa[1], Francisco Marques-Moros[2], Mohsen Rahmani[3], Worawut Khunsin[4], Stefan A. Maier[4,5,6,], Pablo Albella[1,*] and Josep Canet-Ferrer[2,#]

[1] Group of Optics, Department of Applied Physics, University of Cantabria, 39005 Santander, Spain

2 Institute of Molecular Science (ICMol), Universitat de Valencia, 46980, Paterna, Spain.

[3] Advanced Optics & Photonics Laboratory, Department of Engineering, School of Science & Technology, Nottingham Trent University, Nottingham, UK

[4] Department of Physics, Imperial College London, South Kensington Campus, London, SW7 2AZ, United Kingdom

[5] Chair in Hybrid Nanosystems, Nanoinstitute Munich, Faculty of Physics, LMU Munich, Germany

[6] School of Physics and Astronomy, Monash University, Clayton, Victoria 3800, Australia

Corresponding Authors:

*Pablo Albella, email: pablo.albella@unican.es

#Josep Canet-Ferrer, email: jose.canet-ferrer@uv.es

## Abstract

Control over surface temperature is of paramount importance in optoelectronics, photocatalysis and biosensing applications, among others. Thermoplasmonic approaches have demonstrated unrivalled performance for controlling surface temperature, in terms of spatial resolution and thermal amplitude. Most efforts have been done on optimizing the temperature gradient and developing nanoscale temperature patterns. Otherwise, a temperature gradient-free thermoplasmonic surface will enable further functionalities. For example, a constant temperature area of a few square microns can be employed to tune the gap of a 2D crystal or to control transition in a phase change material without the need of local resistive heaters and external electronics. In this work, we present a thermoplasmonic platform conceived for this purpose. It consists of an array of gold nanoantennas on a "silicon on insulator" substrate, where the heat is generated by a laser beam focused on the array, diffused over the whole silicon layer and confined in the vertical direction by the insulator underneath. As an advantage to previous approaches, heat diffusion allows fine temperature control at a desired distance from the excitation spot, thus showing the proposed platform as a candidate for hosting and controlling the temperature during the characterization of light-sensitive materials.

**Introduction**

Resistive loss in metals is considered the main drawback of plasmonic nanoantennas[1]. In the field of thermoplasmonics, this limitation is turned into an advantage by using nanoantennas as optically driven thermal sources[2–8]. This enables contactless real-time surface temperature control, which is extremely attractive for applications in photonics, sensing, drug delivery, bio-imaging, photocatalysis or micro-thermophoresis[9]. For example, thermoplasmonic semi-transparent electrodes have been employed to increase thermal gradients, thus improving the response of a thermoelectric solar cell.[10] Nanopatterned thermal hot spots can be exploited by optical lithography methods to beat the diffraction limit.[11] Meanwhile, recent multidisciplinary research demonstrated the benefits of applying plasmonic photothermal therapy in combination with other medical treatments[12].

The outstanding achievements in thermoplasmonics are bound to the growth of the Plasmonics community during last twenty years. Boosted by the interest on describing light-matter interaction at the nanoscale, accurate numerical methods for the description of the electromagnetic response of plasmonic nanoantennas have been developed. In combination with heat diffusion models, numerical methods provide an effective tool for finding the trade-off between the optical and thermal properties [13]. Just to give some figures, the resolution in temperature patterns can reach hundreds of nanometers limited by the separation between heat sources (e.g. period in nanoantenna array).[14] Under typical experimental conditions the patterning can exhibit thermal gradients on the order of 50 (K·µm$^{-1}$),[15] with numerical simulations foreseen values up to 1000 (K·µm$^{-1}$) under special optical pumping conditions. But there is still room for improving thermoplasmonic applications.

Great efforts have been made on the generation of sub-diffraction limited thermal patterns, however, a simple approach to homogeneous temperature control over large areas is still missing and only achieved by tailoring the excitation beam. Shaping the excitation spot, employing spatial light modulators, or beam steering techniques.[16] All these approaches result rather unfriendly if compared to a regular focused beam excitation. Another characteristic to improve is the method to monitor the surface temperature. In general, photoluminescence or fluorescence imaging are employed, however the accuracy of these techniques can be significantly affected under certain experimental conditions, and importantly, those methods prevent the use of light sensitive materials.

In this work, we propose a wide-area gradient-free thermoplasmonic metasurface controlled by means of a focused laser beam for applications on surface temperature control. The heat sources are metallic nanoantennas distributed on three quadrants of a wide silicon pad optimized for diffusing heat towards the clear quadrant, and thus generating a constant temperature surface. This way, the temperature control is extended at a distance from the excitation spot, enabling the use of photosensitive materials. The thermoplasmonic performance is characterized using a double-spot confocal setup for the simultaneous heat generation and temperature measurement by Raman thermometry. We demonstrate a thermal contrast close to 100 K under excitation powers as low as 25 µW. The resulting thermoplasmonic platform could be optimized for hosting and controlling remotely the temperature in light-sensitive materials e.g. 2D crystals, spin crossover nanoparticles or molecular compounds.

**Experimental Section**

*Numerical Simulations*

Periodic metasurfaces composed by hybrid metallic-dielectric unit cells can be used as heat sources under optical pumping. It is well known that metallic nanoantennas show high resistive loss with the consequent light-to-heat conversion. The conversion is more efficient for incident wavelengths close to localized surface plasmon resonances (LSPR) which can be optimized with a correct selection of the cell material, shape and size. The heat generated by photon absorption at the LSPR is diffused along the nanostructure and its surrounding which is usually optimized to generate a temperature gradient. The resulting spatial distribution of the temperature can be foreseen by solving the heat transfer equation which for the adequate boundary conditions reads,

$$\rho(\boldsymbol{r})c(\boldsymbol{r})\frac{\partial T(\boldsymbol{r},t)}{\partial t} = \nabla k(\boldsymbol{r})\nabla T(\boldsymbol{r},t) + Q(\boldsymbol{r},t) \qquad (1)$$

being $r$ and $t$ the position and time, $T(r,t)$ the local temperature and $\rho(r)$, $c(r)$ and $k(r)$ are the mass density, specific heat and thermal conductivity. $Q(r,t)$ is the energy source coming from light absorption.

Maxwell equations must be solved first to estimate the heating power, i.e., the resistive loss. In most cases, numerical methods must be applied. In our calculations, the whole process of light absorption and further heat transfer has been accounted by means of Finite Difference Time Domain (FDTD) and Finite Element Method (FEM) simulations. For an easy implementation and reliability of the solution, we have chosen Lumerical FDTD and COMSOL Multiphysics 5.6, software which provide integrated solid methods to solve partial differential equations.

All the thermoplasmonic calculations were performed in two different steps, as usual [17–19]. A wavelength-controlled free tetrahedral mesh was used to ensure a high-quality geometrical modelling with an appropriate element density and reliable curvatures. First, we have calculated the resistive losses by solving the electromagnetic part of the problem (RF Module) under linear polarized illumination, thus obtaining the heat source ($Q$).

In the second step, we used a heat flux node across the outer frontiers defined from the heat transfer coefficient to consider heat dissipation. This coefficient was estimated as usual by dividing the thermal conductivity of the fluid by a length scale. All thermal parameters involved in our calculations (density, specific heat and thermal conductivity) were taken from the COMSOL Multiphysics material database. We used Lumerical FDTD to double check our results and eventually analyse spectral reflectance and transmission of the devices under study. The FDTD simulations were carried out by considering a periodic unit cell illuminated at normal incidence with a linearly polarized plane wave. The mesh consisted of a rectangular discretization dense enough to ensure more than fifteen elements per wavelength. All the optical properties were taken from the Lumerical database.

*Fabrication*

Arrays of monomer and dimer nanoantennas were fabricated by electron beam lithography on a backside polished silicon-on-insulator (SoI) substrate, a commercial substrate widely used in integrated photonics. The SoI is composed of a thin silicon film on top of an oxide layer. In particular thickness of top silicon was 220 nm and silica layer underneath 3.5 µm. The fabrication was done via etch-down approach, starting by depositing 10 nm silica and 40 nm Au on the top of SoI wafer. Subsequently, the substrate was coated with negative resist, followed by EBL patterning and developing. Finally, lithography patterns were used as a mask and transferred to the substrate via ion-milling.

*Optical Characterization*

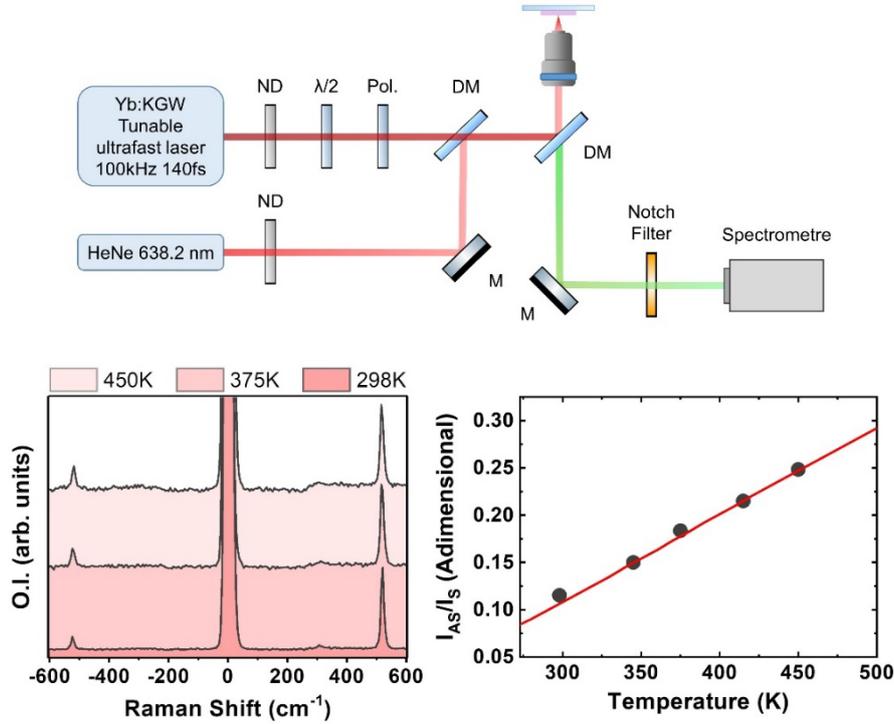

**Figure 1.** (a) Experimental set-up. (b) Raman spectra on silicon at different temperatures. (c) Ratio between the anti-Stokes and the Stokes signal at different temperatures (ASSR). The black scatters correspond to the experimental data which are fitted to Eq. 2 (red line).

The thermoplasmonic effect was characterized by means of an in-house constructed double-spot confocal microscope. The first spot was provided by a HeNe laser at 633 nm wavelength and the second one by a femtosecond pulsed laser at 830 nm. The optical paths were aligned to focus on different points of the sample without any overlap. The backscattering of the red laser was analysed by means of a spectrophotometer to monitor the Raman peaks of Silicon while the infrared laser (only used to excite the nanoantennas) was filtered, see the scheme in Fig. 1a. The calibration of the setup was carried out by measuring the Raman signal of silicon at different temperatures. The surface temperature could be monitored by comparison between anti-Stokes and the Stokes ratio (ASSR), see three representative Raman spectra in Fig. 1b. The corresponding ASSRs are plot in Fig. 1c (black scatters) lying on the curve defined by Eq.2 (red line), typically employed for Raman thermometry:[20,21]

$$ASSR = \left[\frac{E_0 + \Delta E_i(T)}{E_0 - \Delta E_i(T)}\right]^4 \exp\left[-\frac{\Delta E_i(T)}{k_B T}\right] \qquad (2)$$

where $E_0$ is the excitation laser energy, $\Delta E_i$ is the phonon energy, $k_B$ is the Boltzmann constant and $T$ the temperature.

The agreement between numerical and experimental data suggests a negligible local heating effect associated to red Raman laser. Furthermore, an experimental error of about 10 K is estimated at room temperature but considerably reduced at higher temperatures. This sensitivity is satisfactory to characterize our devices, however, two points might be improved for those applications requiring higher accuracy. Both are related with the acquisition conditions: i) the excitation energy selected and the dynamic range of the detector and ii) the integration time. Firstly, under red light excitation the Raman signal of silicon is weak and the anti-Stokes peaks can be affected by the signal-to-noise ratio. This limitation can be overcome by employing a higher energy excitation (e.g. close to one order of magnitude using a green laser) but in our experiment we preferred to use the red laser to reduce possible local heating effects, *see Section SI1 for more details*. Secondly, the stronger Stokes signal, and eventually the high intensity of the laser peak, limits the integration time. Employing monochannel detectors our set-up could be improved to discriminate brighter undesired peaks.

**Results and discussion**

Our approach for a gradient-free thermoplasmonic platform is based on the combination of materials with opposite thermal conductivities. A high thermal conductivity material is used as active layer to homogenize the temperature. Below the active layer, a low thermal conductivity material is required to reduce the heat flow towards the substrate.[22] The potential of this combination is illustrated in Fig. 2 by means of numerical simulations of three simple approaches. The first case consists of a rectangular array of gold nanoantennas located on a low thermal conductivity material (see Fig. 2a), in particular silica ($k = 1.1$ W/mK[23]). Identical nanoantennas are deposited on a high thermal conductivity material, in this case silicon ($k = 156$ W/mK[24]). The third structure consists of a combination of the previous materials in a double layer. In all cases, the array is made of nanodisks with 100 nm diameter and 40 nm height while the period of the lattice is set to 350 nm. A typical incident power density of 1 mW/µm$^2$ ($10^9$ W/m$^2$) is considered in the numerical simulations[25,26].

Important quantitative differences are found in the thermoplasmonic response of these three devices. The optical spectrum of the nanoantenna array on silica presents a reflection peak around λ = 600 nm (see Fig. 2b). On silicon, this peak is broadened and redshifted to λ = 1150 nm due to the higher refractive index of the substrate, see Section SI2. However, when the nanoantenna array interacts with the optical resonances of the double layer, this one drives a delocalization of the induced field. In this situation, we can identify a set of peaks allowing an efficient light-to-heat conversion, see Fig. 2b. In Fig. 2c, the temperature increase is plotted for different input power densities considering excitation wavelengths λ = 600, 1150 and 830 nm, for the silica, silicon and double layer case, respectively. In those ideal structures, the temperature varies linearly with the excitation power, however, the temperature increase in the silicon substrate is negligible compared to the thermoplasmonic response of the other devices.

Figures 2d and 2e correspond to the mappings of the electric field enhancement at the resonance wavelength of the nanoantenna array, i.e. λ = 600 and 1150 nm for SiO$_2$ and silicon respectively. In both cases, the electric field is strongly confined around the nanoantenna. As a difference, on the multilayer (Fig. 2f) the electric field distribution presents an interference pattern. It is worth noting that despite the homogeneous field distribution, the resistive optical loss occurs mainly at the nanoantennas as non-relevant temperature contrast is found in the same double layer without nanoantennas, see *Section SI3*. As illustrated in Figs. 2g-2i, the result is a temperature gradient-free volume of silicon, where the multilayer exploits the capabilities of the silica to confine the heat flow while the high thermal conductivity of silicon allows a homogeneous temperature on the top surface.

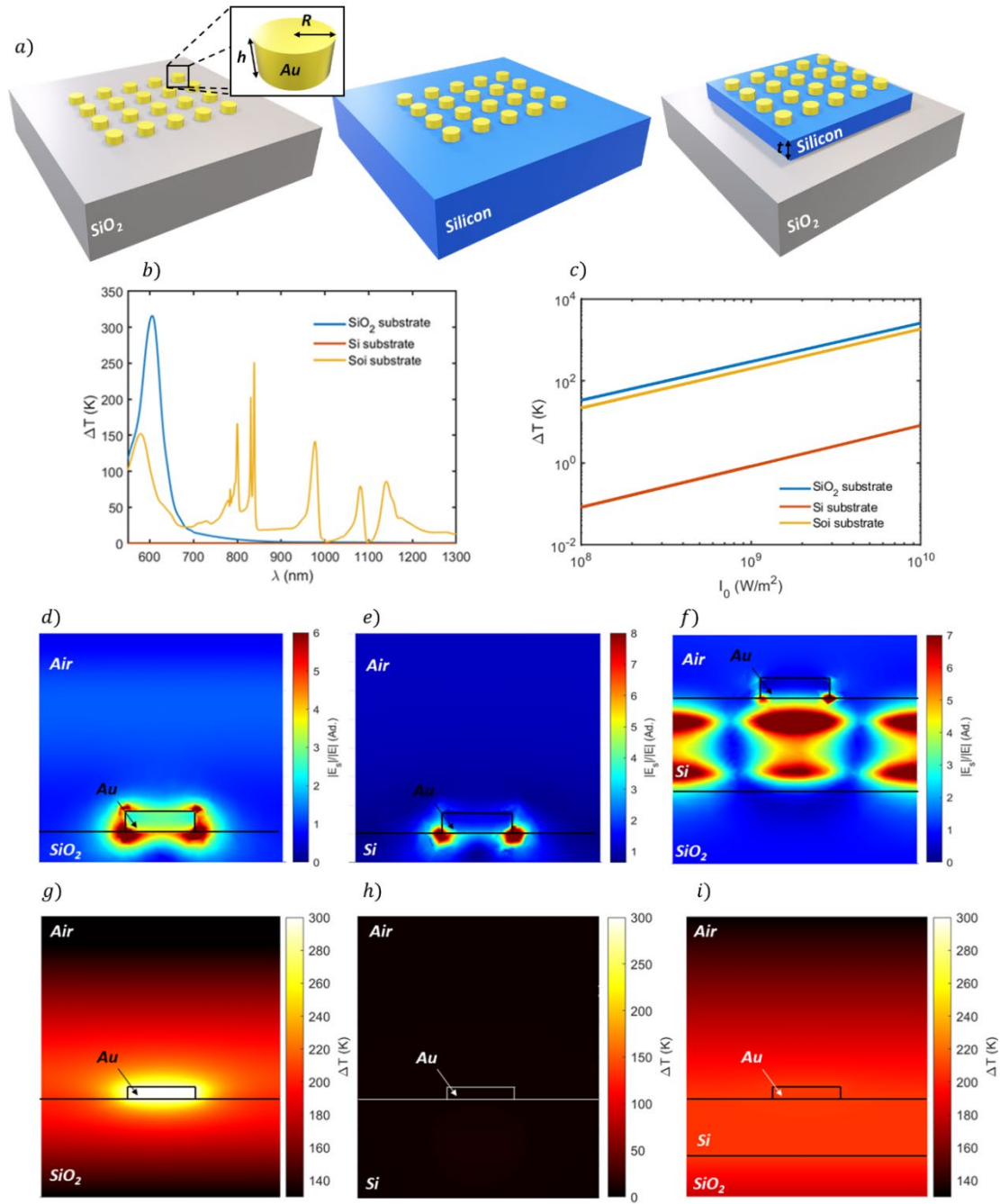

**Figure 2.** Numerical simulations of temperature gradients on three representative cases. (a) scheme of nanoantenna arrays on silica, silicon and a double layer: silicon on silica substrate. Notice the disk parameters, height $h$ and radius $R$, are identical for the three devices and clearly lower than the thickness of the silicon film in the multilayer, $t$. (b) Thermal spectral response of the structures under constant excitation power of 1 mW/µm². (c) Temperature increment as a function of the incident power density. (d), (e) and (f) are the electric field enhancements for the three studied structures excited at 600, 1150, and 830 nm, accordingly. (g), (h) and (i) correspond to the respective temperature mapping.

With these simple devices we have demonstrated the importance of the substrate thermal conductivity. Next, we are going to focus on the role played by the plasmonic array. A deeper view of the above examples reveals details that prevent the proper comparison among the three systems. For example, the resonances on the silicon substrates are redshifted quite below the silicon bandgap. Together other pros and cons, this also raises the issue of the silicon absorption in the SoI substrate. This fact is discussed in *Section SI4*, where we concluded that the proper comparison must be carried out between devices fabricated on different substrates but operating under similar experimental conditions. In *Section SI4* we have also discussed how to find the proper comparison conditions by including a thin silica film between the gold nano-antennas and the corresponding substrate. As a result, resonances of devices on silicon substrates are blueshifted and we extracted valuable information to fabricate a series of reference samples for comparing the experimental results.

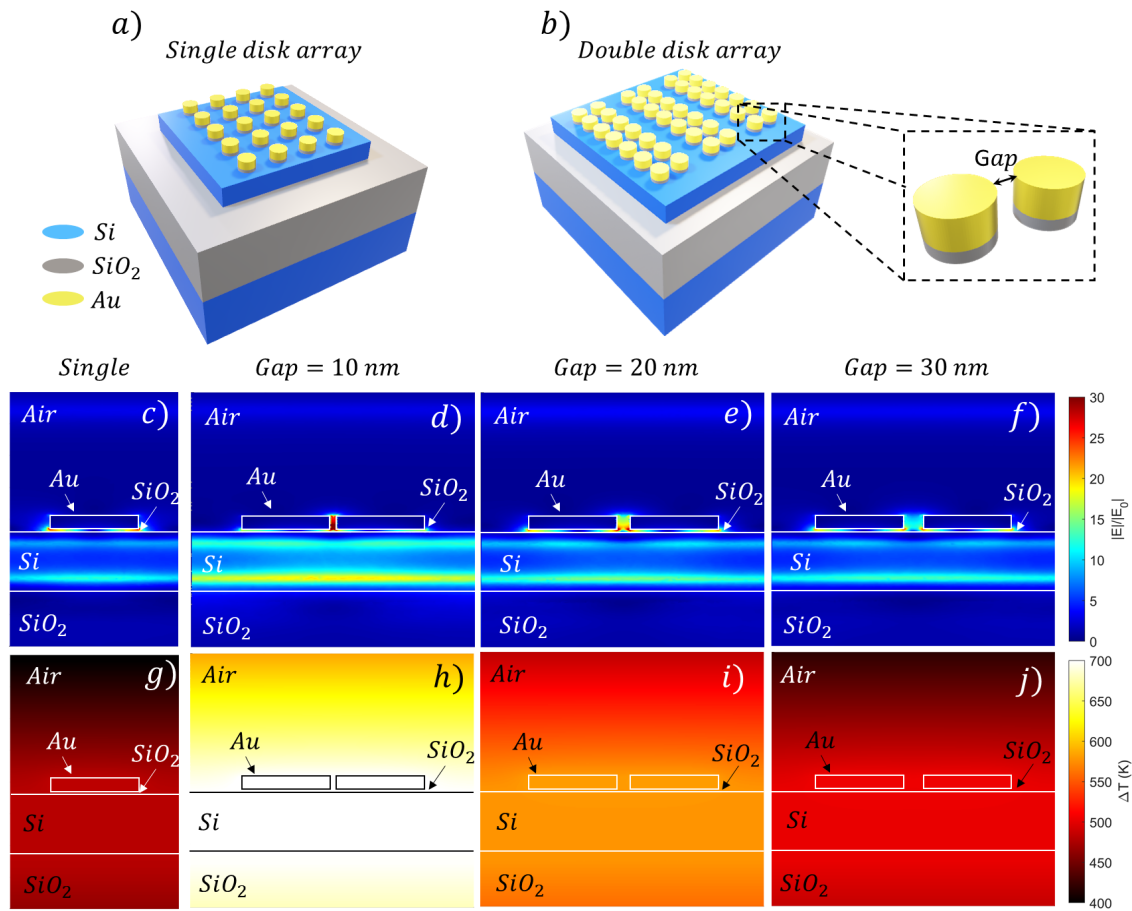

**Figure 3.** Scheme of the gradient-free thermoplasmonic platform built of a single disk array defined by $P_x = P_y = 250$ nm (a) and a double disk array with $P_x = 420$ nm and $P_y = 250$ nm (b). The disks have 60 nm radius and 40 nm height. Numerical simulation of the array electric field enhancement on a SoI substrate excited at 835 nm wavelength for the single array (c) and a the double one for gaps of 10 nm (d), 20 nm (e) and 30 nm (f). The corresponding thermal mapping is simulated in (g), (h), (i) and (j).

In addition, the optimization of our thermoplasmonic platform is carried out considering different kinds of plasmonic arrays. In particular, the comparison among disks and disk dimers with different gap distances (10, 20 and 30 nm) gives a clear picture of the role played by the plasmonic array. For simplicity the size of the single disks and disk dimmers is the same; i.e. diameter of 120 nm, height of 40 nm and separated from the top silicon surface with a ten-nanometre silica buffer (see Fig. 3a). Figure 3 shows the thermoplasmonic response of the four different systems. The electric near-field (Fig. 3c-f) and temperature maps (Fig. 3 g-j) correspond to the case of normal incident light λ = 835 nm. A clear electric field enhancement is present around the nanostructures, coming from a set of narrow optical resonances between 800 and 850 nm, *see Section SI4*. The resulting temperature map foresees a homogeneous temperature with a negligible gradient on the silicon. Importantly maximum temperature achieved is decreased as the dimmer gap is increased, dropping from 700 K (for gaps of 10 nm) to 500 K (for gaps of 30 nm). The strong dependence on gap is another proof of the important contribution of the nanoantenna to the resistive loss in form of an eventual absorption of the silicon top layer. The high temperature increment is explained in terms of the incident power density (1 mW/$\mu m^2$) and the disks size[27]. In any case, the desired thermal gradient-free effect is achieved in all the samples.

Our numerical findings are experimentaly demonstrated by comparison of four different nanostructures, namely: DA (disk array), DDA10 (disk dimer array, nominal gap of 10 nm), DDA20 (disk dimer array, nominal gap of 20 nm gap), DDA30 (disk dimer array, nominal gap of 30 nm ). In practice the actual gap is very hard to estimate, see Fig. 4a. In every platform, the nominal size of the gold nanodisks is 120 nm diameter and 40 nm height. In the case of DDA10, DDA20 and DDA30, the arrays where fabricated with periods of Px = 420 nm (longitudinal to the dimer axis) and Py = 250 nm (transversal to the dimer axis) while in the case of DA, the periods are Px = Py = 250 nm. The silicon layer was patterned to form micrometric square pads (4x4 $\mu m^2$) and the array was patterned leaving a blank region at the top right corner which have been used for measuring the pad temperature with the red Raman laser.

The scattering of the micrometric silicon pad in combination with the reflectivity of the multilayer prevents to characterize the plasmonic resonance of the arrays by optical methods (*see section SI7*). High resistivity silicon (<20 KOhms/$cm^2$) was used as reference to demonstrate that the analyzed array of nanodisks has negligible thermal effect on regular silicon substrates. The performance of the devices was studied by focusing red laser on the clear area of the silicon pad to monitor the Raman signal with a constant irradiation power of 0.25-1 mW. In parallel to the red optical path, the heating infrared laser is aligned to illuminate the centre of the array exhibiting a slightly defocused spot of around 2 $\mu$m diameter. With the heating laser off, we observe a clear Stokes signal accompanied by a weak anti-Stokes driving to ASSR close to 0.11, as expected for room temperature measurements.

The dependence of ASSR on the infrared laser power is shown in Fig. 4b. The Raman spectra is accompanied by a fluorescence background that we attribute to the residua of the nanofabrication process, *see section SI6*. Despite this background, the narrow Raman peaks can be easily fitted, and we can clearly observe dependence of the ASSR on the excitation power. Indeed, the four arrays exhibited high light-to-heat conversion capabilities. As a drawback, we observed a low damage threshold, in some samples around 75 $\mu$W of irradiated power. For this reason, the samples have been characterized applying a maximum average power of 25 $\mu$W. Under this illumination conditions DA sample reached an important temperature increase with ASSR = 0.21, see Fig. 4a. Qualitatively, the rest of the devices present the same behaviour with a decrease in performance; ASSR = 0.19, 0.18 and 14 for DDA10, DDA20 and DDA30 respectively. The higher heating efficiency observed in the DA is attributed to the larger number of hotspots in this array, i.e. 180 hot spots (one each disk) compared to 114 in the other arrays (one each dimmer). At the single device level, higher performance is expected for dimmers compared to disks. Hence, our result points to the importance of the filling factor in the case of arrays and ensembles.

Consistently with numerical simulations, the higher heating efficiency is found in the lowest gap design (DA10). In Fig. 4c, the ASSR data are translated into temperature using Eq. 2 (t*he details are described in section SI1)*. We found maximum temperatures of $T$ = 409, 384, 378 and 329 K, for DA, DDA20, DDA25, DDA30 respectively, which means a thermal contrast for $\Delta T$ = 112, 87, 81 and 32 K with respect to the room temperature, $T_{RT}$ = 297 K. As above mentioned we estimated an experimental error of about 10 K in the surface temperature measurement. Analogous measurements have been carried out in reference samples consisting of silicon pads without nanoantennas. The results strongly support the important role of the array in the thermal response of our thermoplasmonic platforms, see SI5 for further details.

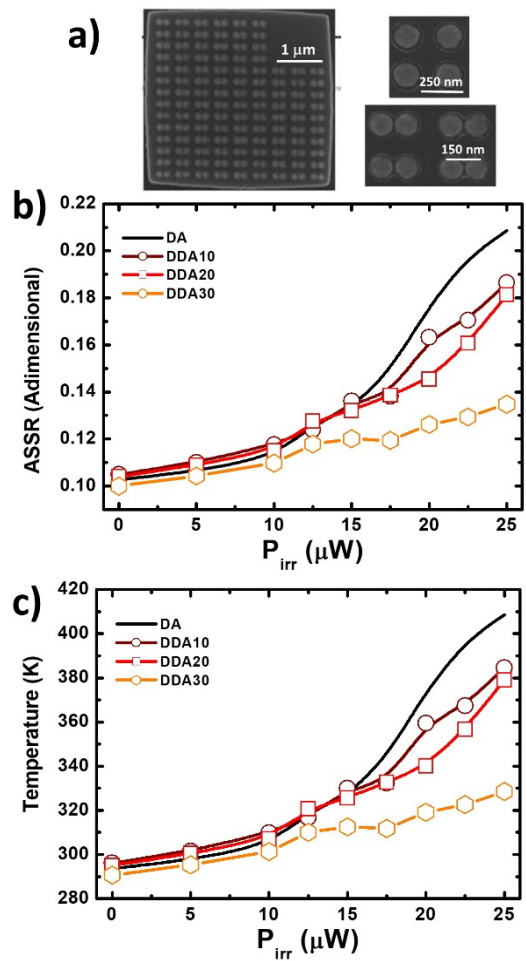

**Figure 4.** (a) SEM images of the fabricated systems. (b)-(c) ASSR and temperature evolution on the irradiated power.

It is also worth mentioning that Temperature vs Power curve exhibits a non-linear behaviour. We attribute this unexpected behaviour to the finite dimensions of our experimental system. The fact is that our numerical simulations consider an extended and periodical illumination source while in the experiment the arrays are excited using a finite Gaussian beam. In these conditions, the behaviour of the system will depend on the range of excitation power. At very low powers, the heat is easily diffused generating a small gradient around the excitation spot. As the excitation power is increased, the gradient approaches the limits of the silicon pad preventing further diffusion. Then, at low and medium powers the heat is confined into silicon pad and the thermal waves are reflected smoothing the gradient over the material. At this moment, the system reaches the linear regime with temperature increasing proportional to the excitation power.

## Conclusions

We report the numerical simulation and the corresponding experimental realization of a thermoplasmonic platform for application as gradient-free temperature control platform. Our proposal exploits thermal conductivity contrast between silicon (used as thermal diffuser) and silica (used to confine the heat on a silicon layer) on a SoI substrate. The thermal energy is generated by an infrared laser beam focused on an array of gold nanoantennas. By means of numerical simulations we illustrate the impact of the different layers on the electromagnetic and thermal response of the system. Numerical results foresee highly efficient light-to-thermal energy conversion and diffusion over a homogeneous temperature area. These results are experimentally demonstrated with the fabrication of different gold arrays on a SoI substrate. Temperature contrasts up to $\Delta T \approx 100$ K have been measured by means of Raman Thermometry under low excitation power (<25 μW). Moreover, the temperature control is extended up to certain distance from the excitation spot. This enables temperature control over non-irradiated areas as required for developing local heating experiments on e.g. 2D crystals, spin crossover nanoparticles or molecular compounds. We believe that the results and conclusions of this study can open a new path towards the development of powerful tools in the area of nanophotonics and thermoplasmonics.

## Acknowledgements

We acknowledge the financial support from the Generalitat Valenciana via the GentT program (CIDEGENT/2018/005); the Spanish MICINN (PGC2018-096649-B-I, PID2020-117152RB-I00, Excellence Unit María de Maeztu CEX2019-000919-M. J. G-C and P.A thank MICINN for his FPI grant and Ramon y Cajal Fellowship (Grant No. RYC-2016-20831), respectively. W. K. acknowledges Marie Sklodowska Curie Actions fellowship received from the European Union's Seventh Framework Programme under REA grant agreement (no. 623424). M. R. acknowledges support from the Royal Society and the Wolfson Foundation. S.A. Maier acknowledges support from the EPSRC and the Lee-Lucas Chair in Physics. All authors appreciate the use of nano-fabrication equipment at Imperial College London.

## Author contributions

P. A. and J.C.-F. conceived the original idea and coordinated the research activities. The numerical simulations were carried out by J. G.-C. under supervision of P. A. The nanofabrication process and preparation of the samples was carried out by M. R. W. K. and F. M.-M. characterized the thermoplasmonic platform and the reference samples under supervision of J. C.-F. All the authors contributed to the data processing and manuscript writing.

## Competing interests

All the authors declare no competing interests.

**Supplementary information**

As supplementary information we include details about the calibration of the Raman thermometer. Numerical spectra describing the role of the substrate and the nanoantenna arrays, numerical optimization of the thermoplasmonic platform, our results on reference samples and some representative peaks of the deconvolution fittings used to calculate the integrated intensity of the stokes and anti-Stokes signal.

# Supplementary Information: Gradient-free temperature control over micrometric areas for thermoplasmonic applications in micro-and nano-devices.

Javier González-Colsa[1], Francisco Marques-Moros[2], Mohsen Rahmani[3], Worawut Khunsin[4], Stefan A. Maier[4,5,6,], Pablo Albella[1,*] and Josep Canet-Ferrer[2,#]

[1] Group of Optics, Department of Applied Physics, University of Cantabria, 39005 Santander, Spain

2 Institute of Molecular Science (ICMol), Universitat de Valencia, 46980, Paterna, Spain.

[3] Advanced Optics & Photonics Laboratory, Department of Engineering, School of Science & Technology, Nottingham Trent University, Nottingham, UK

[4] Department of Physics, Imperial College London, South Kensington Campus, London, SW7 2AZ, United Kingdom

[5] Chair in Hybrid Nanosystems, Nanoinstitute Munich, Faculty of Physics, LMU Munich, Germany

[6] School of Physics and Astronomy, Monash University, Clayton, Victoria 3800, Australia

## Section SI1. Calibration of the Raman thermometer

There are different methods to measure the temperature of a sample through its Raman signal, such as the intensity ratio between stokes and anti-stokes bands, the peak position, and the linewidth of the bands. Their applicability depends on the sample, the experimental conditions, and the system used. In this way, some limitations can be found depending on the selected method. For example, the peaks position is limited by its sensitivity to changes in the distances between atoms, such as thermoelastically induced strain, causing error in the thermometry measurements. In addition, depending on the resolution of the detector, the amount of error introduced in the thermometry can be noticeable.

Here we used the anti-stokes-to-stokes ratio (ASSR), which arises from differences between phonon population states as the temperature changes. In other words, Raman thermometry based on the ASSR relies entirely on physical mechanisms that change the phonon distribution from their equilibrium (which follows a Bose-Einstein distribution). Thus, the phonon population of the ground state at lower temperatures is larger than that of the excited state and the ASSR yields to zero at very low temperatures. As the temperature increases, the ASSR rises up following the corresponding thermal distribution reaching values around 0.1 in the case so silicon at room temperature.

The ratio of the anti-Stokes over the Stokes Raman peaks comes from the following equation:

$$\frac{I_{AS}}{I_S} = \frac{(V_I + V_v)^4}{(V_I - V_v)^4} e^{\left(\frac{-hV v}{kT}\right)} \quad (1)$$

Where $I_{AS}$ and $I_S$ are respectively the anti-Stokes and Stokes scattering intensities, T is the temperature, h is Planck's constant (6.626·10$^{-34}$ J·s), and k is the Boltzmann's constant (1.38065·10$^{-23}$ J/K), using the International System. Accordingly, the excitation source and the Raman mode frequency come in s$^{-1}$. In our particular case:

For $V_I$, the frequency of the red Raman laser:

$$c = \lambda \cdot V_I \quad where\ \lambda = 632.8\ nm;\ c = 2.9979 \cdot 10^8\ \frac{m}{s} \tag{2}$$

$$V_I = 4.73755 \cdot 10^{14}\ s^{-1}$$

For $V_v$, which is the frequency of the vibrational Raman mode:

$$\tilde{v} = \frac{1}{\lambda} = \frac{V_v}{c} \quad where\ \tilde{v} = 520\ cm^{-1};\ c = 2.9979 \cdot 10^8\ \frac{m}{s} \tag{3}$$

$$V_v = 1.558921 \cdot 10^{13}\ s^{-1}$$

In this way, we get that for the Si and using a 632.8 nm laser:

$$\frac{I_{AS}}{I_S} = 1.30126 e^{(-\frac{748.1556}{T})} \tag{4}$$

In some case, it would be easier to calculate the adimensional coefficient using magnitudes in $cm^{-1}$. Then, laser = 15802.78 cm-1 and the phonon energy = 520 cm$^{-1}$. Also, we can use Boltzmann constant in eV/K, the plank constant in eV*s and the phonon energy in eV. Then,

$$V_v = 64.47 \cdot 10^{-3}\ eV$$

Temperature measurements with this method are known to lose accuracy at very high and very low temperatures. At high temperatures small ratio variations drive to in large temperature changes (Abel et al., 2007) while the weak signal of the anti-stokes at very low temperatures drives to an inaccurate temperature estimation. However, at room and slightly higher temperatures it presents some advantages. Firstly, measuring conditions have a minor impact on the temperature estimation, as a difference to those methods monitoring the phonon energy shift and linewidth broadening, which require high resolution and accurate spectral calibration. Secondly, the ASSR allows to identify laser heating effects, which is a great advantage in this work.

To illustrate this, we have carried out Raman measurements in a bare SoI substrate at four excitation energies at different powers: 473, 532, 633, and 780 nm wavelength between 0.02 to 2 mW excitation power. The Raman signal clearly increases with the excitation energy, being the spectra acquired with the 473 nm laser the most intense signal. In addition, we can foresee higher absorbance at this wavelength. Indeed, we have distinguished heating effects in Figure SI1a. Similar measurements produce negligible heating effects when exciting with red and infrared light. For this reason, despite 473 nm or 532 nm provide better signal-to-noise ratio, we decided to use a HeNe laser for the thermometry measurements in this work.

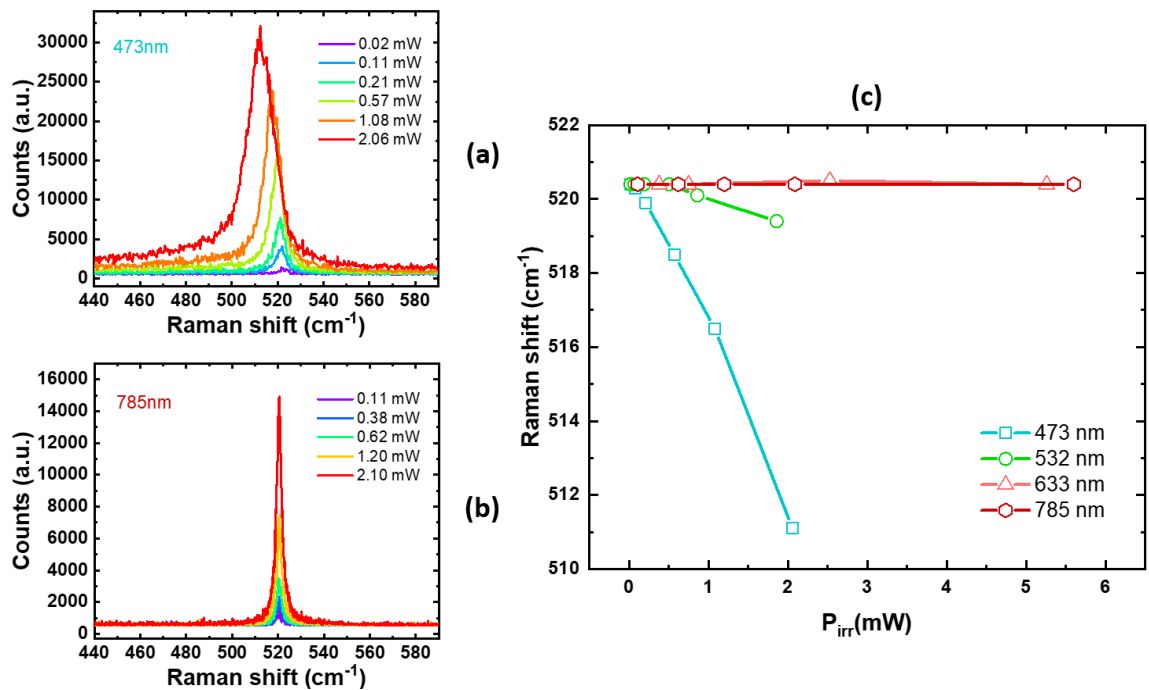

**Figure SI1** (a) First order Si Raman band with 473 nm and (b) 785 nm excitation wavelengths at different irradiation powers. (c) Dependence of the central wavenumber with the irradiation power at 4 different excitation wavelengths. The measurements correspond to a reference SoI sample at 1 second acquisition.

Finally, it is worth mentioning that the phonon peak monitored in this works correspond to the vibrational mode of bulk silicon (diamond structure). The corresponding symmetry allows one first-order Raman active phonon located at the Brillouin zone (BZ) centre (Iatsunskyi et al., 2015). The intensity of this first-order scattering band comes from the optical phonons TO and LO at the centre point Γ of the BZ. This corresponds to the highest intensity Raman band located at $520.0 \pm 1.0$ cm$^{-1}$ (at room temperature), which is shown in Figure 1. In addition to the first-order band, it can be observed the second-order broad band at a higher frequency. This Raman signal is usually located between 900 and 1100 cm$^{-1}$ (950 cm$^{-1}$ in Figure 2). It comes from the scattering of three different transverse optical 2TO phonons: 2TO (X), 2TO (W) and 2TO (L) (Iatsunskyi et al., 2015).

For this second-order scattering, the two phonons taking part must have equal but opposite wavevector in order to fulfil momentum conservation (Meier et al., 2006). Therefore, the strongest scattered signal comes from phonons where the density of states (DOS) is highest (Cardona & Yu, Y., 1996). The intensity of this second order Raman band strongly depends on the scattering geometry due to symmetry reasons, which in our lab we use as calibration and for double checking the quality of our Raman signal.

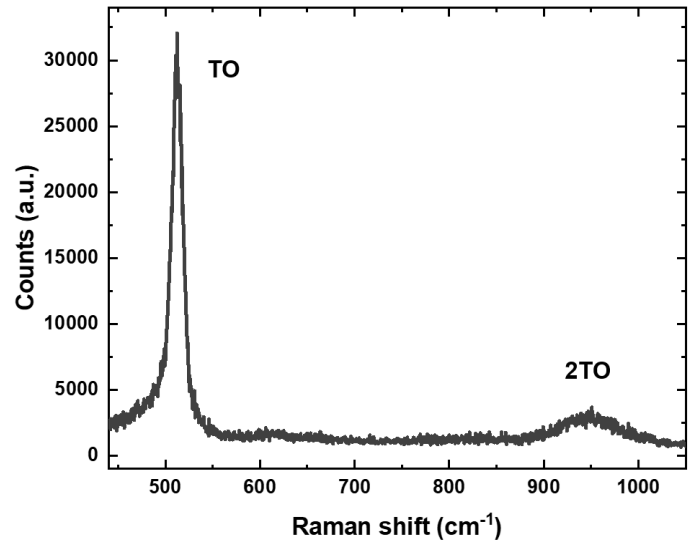

**Figure SI2** Raman spectrum of SoI with 473 nm excitation wavelength showing typical Silicon 520 cm$^{-1}$ LTO and 950 cm$^{-1}$ 2TO optical phonon bands.

## Section SI2: Transmission and reflection spectra of the analysed systems.

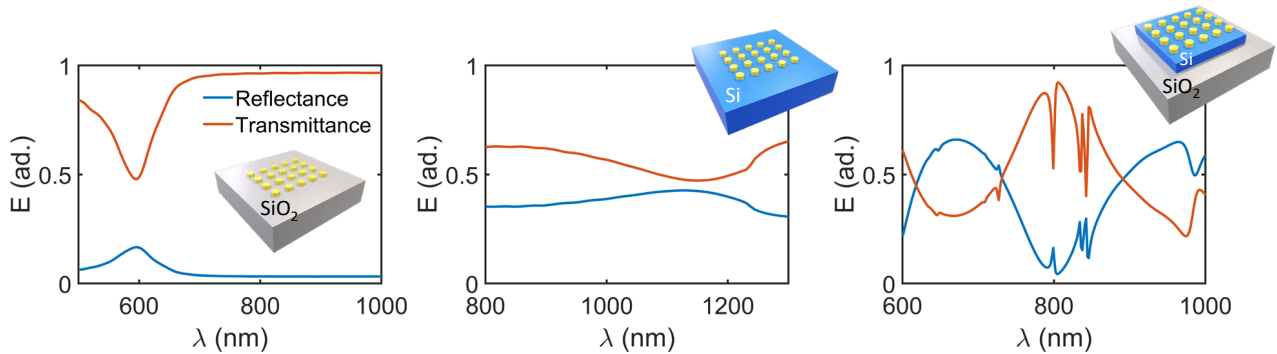

**Figure SI3** Reflectivity and transmittance of a gold disk array on glass (a), silicon (b) and silicon on glass (c) illuminated at normal incidence.

In figure SI3 it can be seen how the substrate material influences the spectral response of the gold nanodisks array. In Figure SI3a, the off-resonance of background reflectivity is negligible and the energy is mostly transmitted as expected. However, there is a remarkable absorption at 600 nm because of the array's plasmonic response. Conversely, when silicon is considered as a substrate, an increment in the background reflectivity can be seen due to the high refractive index of silicon. It can also be noticed that the absorption is dramatically redshifted with respect to the case of the glass substrate and also broadened. The spectral response of a combination of layers is shown in figure SI3c. The background reflectivity and transmission patterns mostly correspond to a Fabry-Perot resonator (multilayer) composed by a 220 nm-thick silicon layer on a semi-infinite silica layer. For the nanodisks array on the multilayer, a sharpened minimum appears in the transmittance pattern corresponding to the plasmonic excitations.

## Section SI3. Study of the multilayer structured substrate.

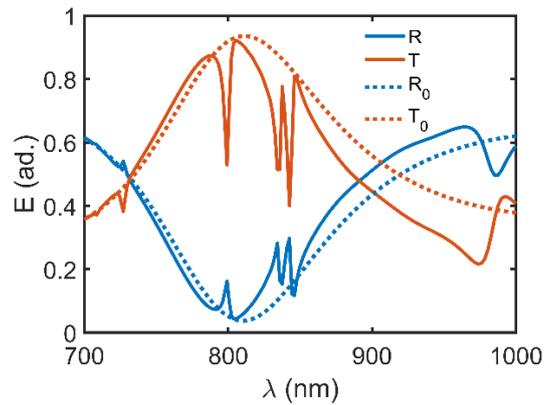

**Figure SI4** Reflectivity and transmittance of a SoI substrate with (solid) and without (dashed) a gold disk array. The parameters of the array and substrate corresponds to the devices in Fig. SI3 of the manuscript.

In figure SI4, the reflectivity and transmittance of a SoI substrate with and without gold nanodisks are compared. The silicon layer has 220 nm thickness and the nanostructures have a period of 350 nm. It can be observed that in absence of metallic nanodisks (dotted line), the multilayer behaves as a Fabry-Perot, showing the characteristic spectra for the mentioned thickness. As a difference, remarkable minima in the transmittance are produced in the presence of plasmonic antennas. The absorption related to these peaks generates resistive loss and drives to local heating. Roughly speaking, the spatial field distribution is determined by the multilayer structure which enhance the electric field below the nanoantenna array as shown in the Fig. SI5.

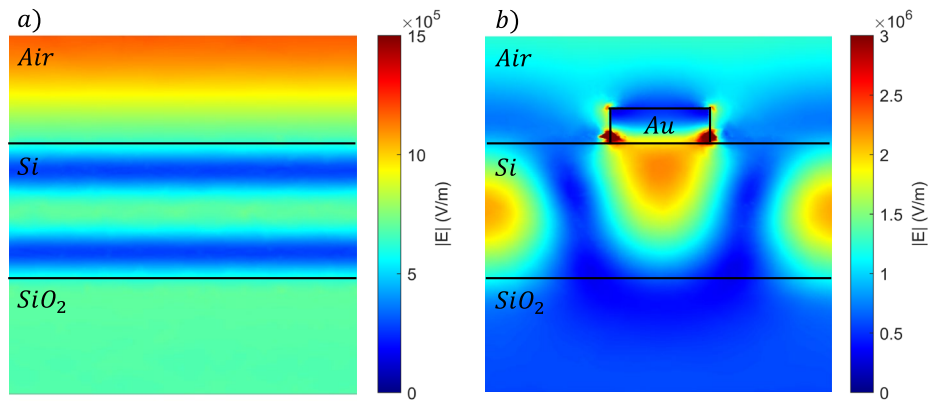

**Figure SI5** Normalized electric field distribution of the devices of Fig. SI4 excited at 830 nm wavelength for a SoI substrate in absence of nanostructures (a) and in presence of nanodisks (b). Notice the different scale bare of the figures.

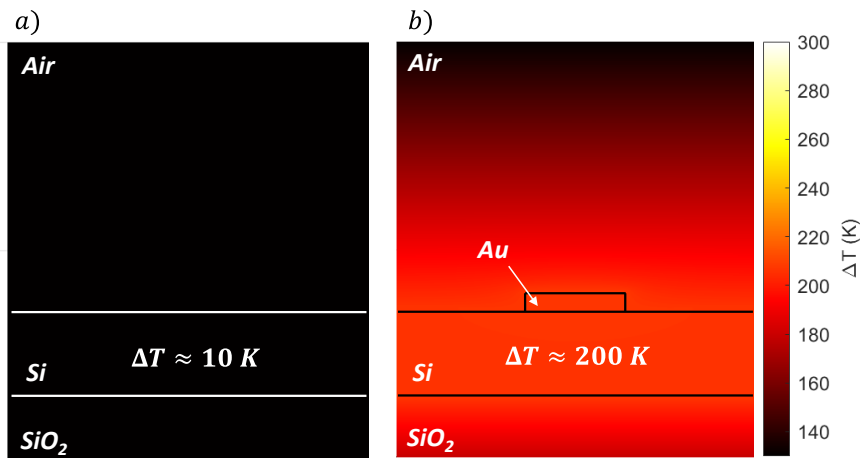

**Figure SI6** Two-dimensional spatial thermal distribution of the Fig. SI4 devices. SoI substrate in absence of nanostructures (a) and in presence of nanodisks (b). Both metasurfaces are illuminated at 830 nm wavelength with a power density of 1 mW/µm².

Figure SI6 show the multilayer spatial temperature distribution for both situations, with and without nanodisks. Despite we can observe certain field enhancement into the slab in Fig. SI6a, the local heating is negligible. This low performance is attributed to the low imaginary permittivity in the range of our study. In addition to increasing the near-field enhancement, the plasmonic nanoantennas provide the resistive loss required for and efficient light-to-heat conversion.

### Section SI4. Influence of the silica buffer on the gradient free metasurface.

As discussed in the manuscript, the comparison among the simple devices in Fig. SI3 gives a first insight about the advantages of combining high and low thermal conductivity materials in thermoplasmonic devices. Experimentally, it was hard to make a comparison among devices fabricated on silicon and SoI operating in similar conditions. The main challenge consisted of obtaining devices fabricated on different substrates but operating at similar wavelength as plasmonic resonances of nanoantennas are strongly redshifted on regular silicon. The most similar operation conditions were achieved including a thin silica film between the nanoantennas and the corresponding substrate (either silicon or SoI). In this case, we could provide more convenient conditions for comparison.

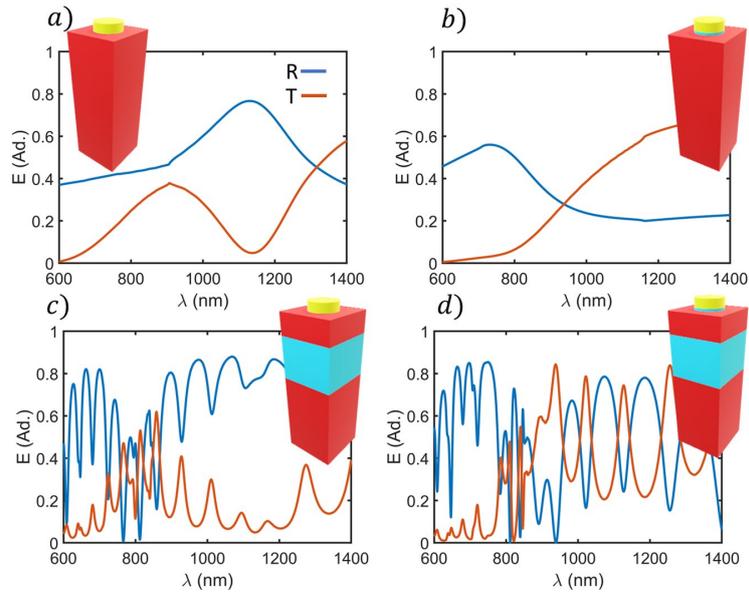

**Figure SI7** Effects of the thin silica film below the nanoantenna.

The highest impact on the optical properties was observed on the devices deposited on regular silicon substrate which resonances are blueshifted above the silicon bandgap. This was important because at some we could question the role of the absorption of silicon. The impact of the silica buffer on the SoI devices is very low, as discussed below, and the other way around for the thermal properties. The thermal performance of the devices fabricated on SoI is clearly reduced by the action of the thin silica film while silicon devices present a very low thermal performance and the presence of the thin silica film between the gold and the substrate was irrelevant.

In the manuscript, we assumed this reduction of performance for the sake of the comparison among SoI and regular silicon devices. In this section, we will demonstrate that the results shown in the manuscript can be considerably improved by removing the silica between the nanoantenna and the top silicon layer of the SoI.

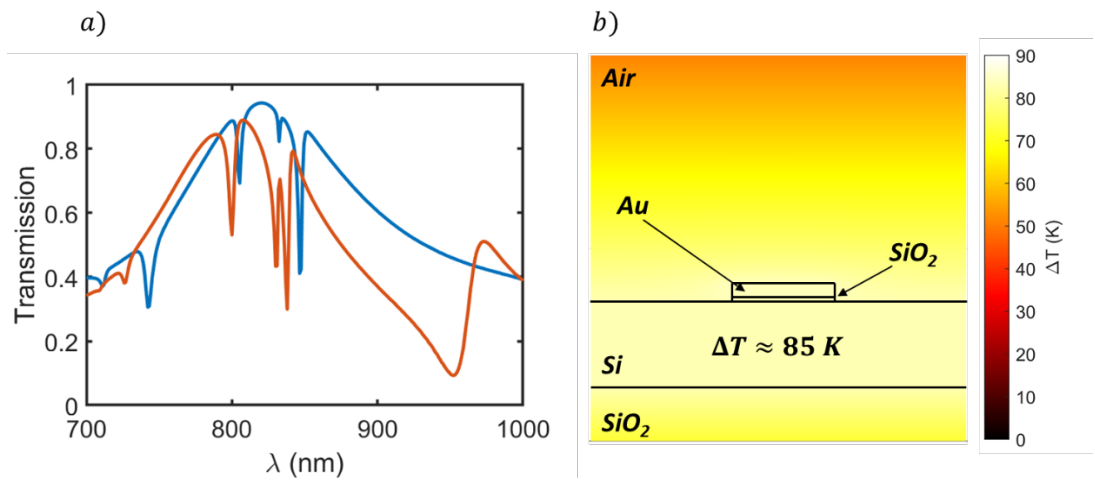

**Figure SI8** a) Transmission of the system with (blue line) and without (orange line) silica buffer. b) Thermal map of the multilayer with silica buffer and gold nanodisks illuminated at a wavelength of 830 nm.

In figure SI8 the behaviour of the multilayer with and without silica buffer is shown. In Figure SI8a it can be seen that the transmission spectrum is blueshifted when the silica buffer is added, and the minima are also diminished. This increment in transmission can be seen as an absorption reduction, thus, reducing the metasurface thermal capabilities as shown in figure SI8b. However, the silicon layer still acts as a thermal conductor in contrast with air and silica distributing the heat which results in a homogeneous temperature distribution as required for our pursued applications.

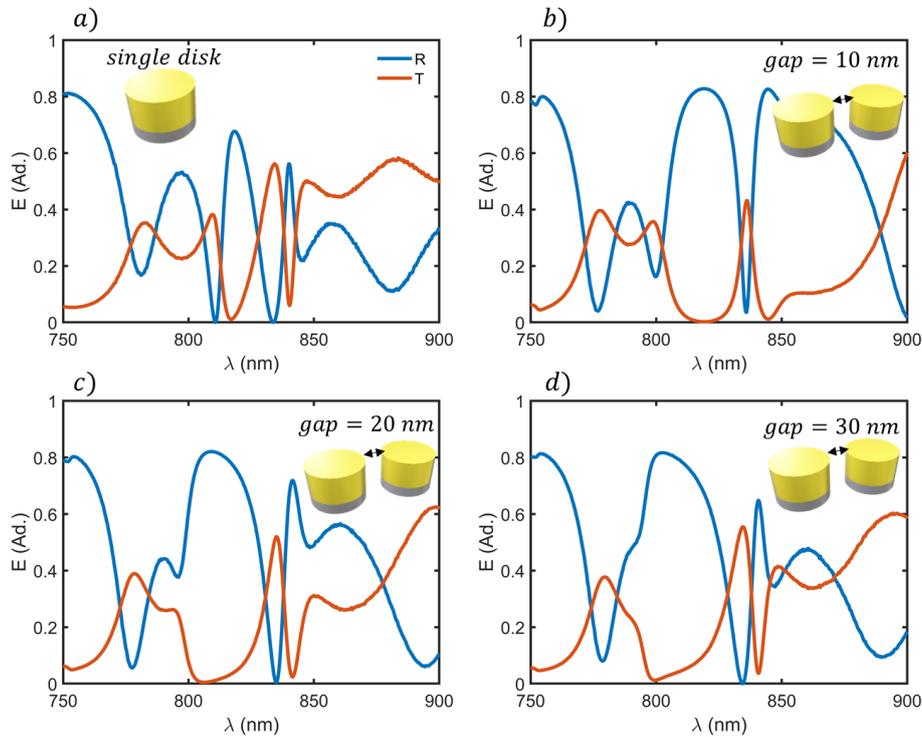

**Figure SI9** Reflection and transmission of the system for the single disk array (*a*), and the dimer array for gaps of 10 nm (b), 20 nm (c) and 30 nm (d) on a SoI substrate. The disk structure consists of a gold disk with 60 nm radius and 40 nm height on a 10 nm-thick silica layer. We can find important resonances between 800 to 850 nm in any case.

Figure SI9 shows the reflectance of the nanostructured systems on SoI for different geometrical parameters, i.e., the type of the disk array and the gap distance between disks. It can be seen how the shape of the background spectra coming from an interference layer are modified by the presence of the nanostructured arrays. Focusing on the wavelength of interest, it is noticed that around $\lambda = 835$ nm, a minimum in reflectivity takes place for the simple disk array (figure 9a) which corresponds to a maximum in transmission. The minimum also appears for the rest of cases. This mode is related to the layer interference that amplify the resistive losses in the gold arrays. As can be seen in figures SI9b-d, this mode is slightly shifted, thus presenting low spectral dependence on the gap distance. Therefore, although the spectral position of the resonance is almost invariant upon gap changes, a thermal contrast is expected since the magnitude of the peaks is remarkably modified.

## Section SI5. Reference samples for the thermoplasmonic characterization

In this section we compare the disk array thermoplasmonic platform with a reference sample and two control samples. The comparison is shown in Fig. SI10. The reference sample correspond to an identical array fabricated on regular silicon (yellow line). The control samples are two different silicon pads fabricated on the same substrate than the disk array but without nanoantennas. One of the pads has been left in blank and contains (red line) while the other one (wine color line) has been covered with a thin double layer film of silica (10 nm) and gold (40 nm). Negligible heating effects (within the experimental error) can be attributed to reference sample and to the silicon control pad.

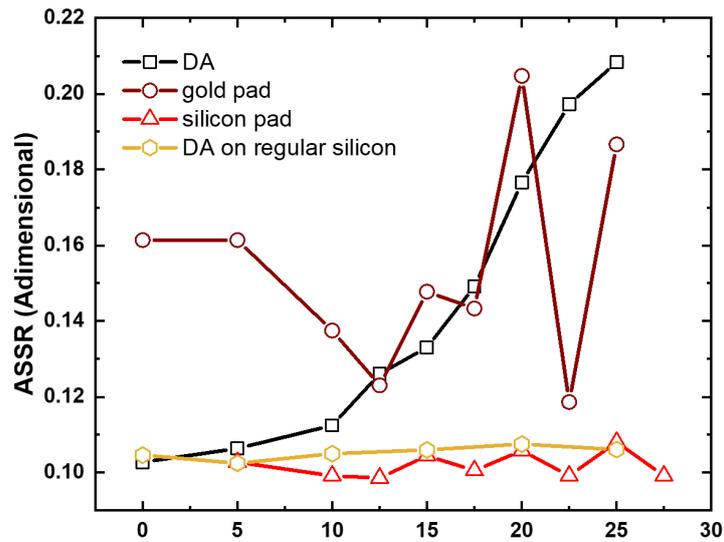

**Figure SI10** Thermoplasmonic response of the disk array sample (in black) compared to re reference sample (in yellow) and the control samples (silicon pad in red and gold pad in wine color)

The gold pad corresponds to a Fabry-Perot cavity with a metallic mirror, so we could expect some heating effects. For some wavelength different to 830 nm these effects could be even comparable to the case of the DA relevant. However, if we split the ASSR in Fig. SI11 in the corresponding anti-Stokes and Stokes intensity we will understand that measurements corresponding to the gold pad are inaccurate. See Section 6 for more details.

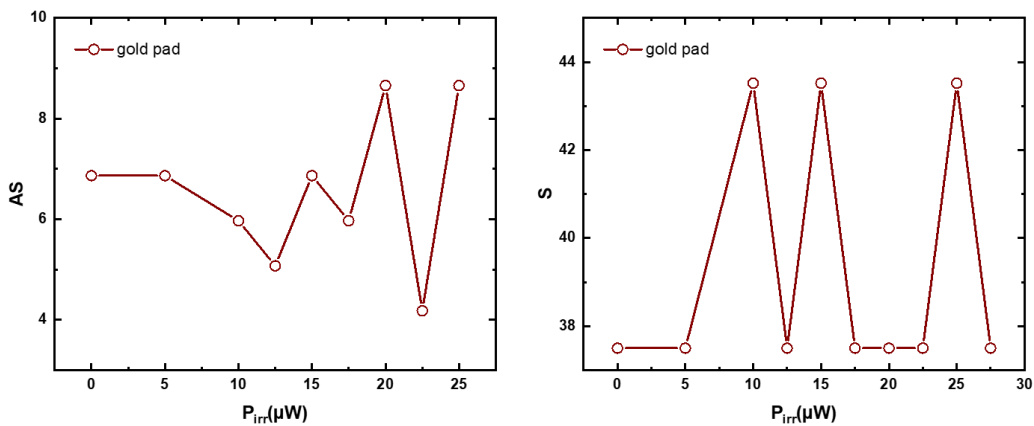

**Figure SI11** stokes (a) and anti-stokes (b) signal evolution of the infrared excitation power on the gold control pad.

### Section SI6. Deconvolution of Raman peaks

In this section we show three representative Raman spectra. The spectrum in Fig. SI12a is acquired in a regular silicon sample, so the signal is clean, and the corresponding temperature estimation will be determined by the sensitivity of the detector. Figure SI13 corresponds to measurements carried out in the clear area of a thermoplasmonic platform. As mentioned in the manuscript, important background is found that we attribute to fluorescence arising from residua due to the nanofabrication process. Despite the background, the narrow Raman peaks can be easily fitted driving a rather accurate estimation of the surface temperature, but in this case limited by the peak deconvolution error. Finally, Fig. SI14 corresponds to gold reference pad (where the plasmonic array is substituted by a thin silica and gold double layer). In this case, the Raman peaks are very weak, and the signal-to-noise ratio drives to an inaccurate temperature estimation due to an unbalanced overestimation of the integrated intensity of the Raman peaks.

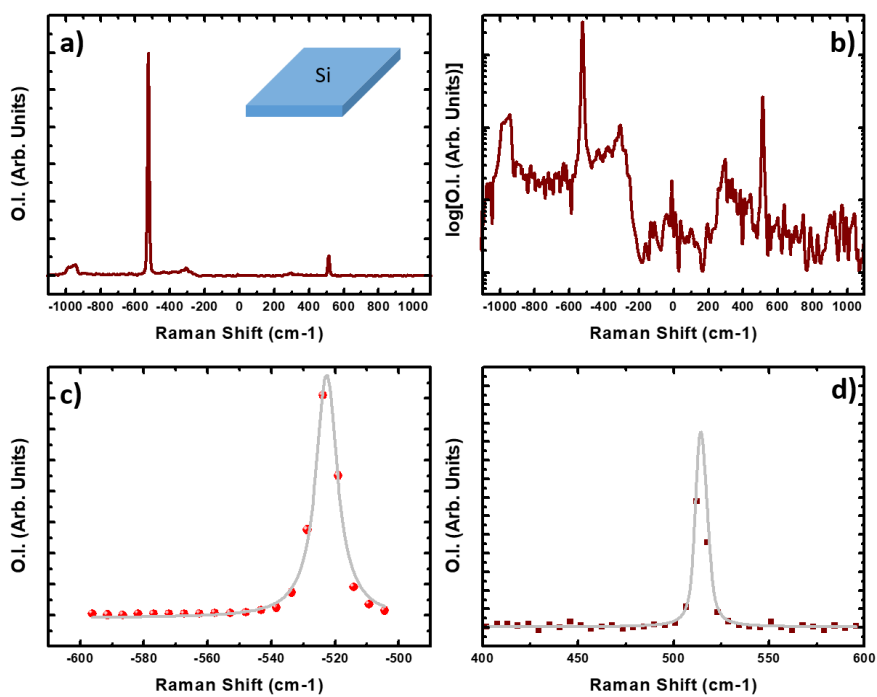

**Figure SI12**. (a) and (b) Raman spectra acquired on a flat silicon sample in arbitrary units and logarithmic arbitrary units respectively. (c) and (d) Lorentzian fitting to Stokes and antiStokes, respectively.

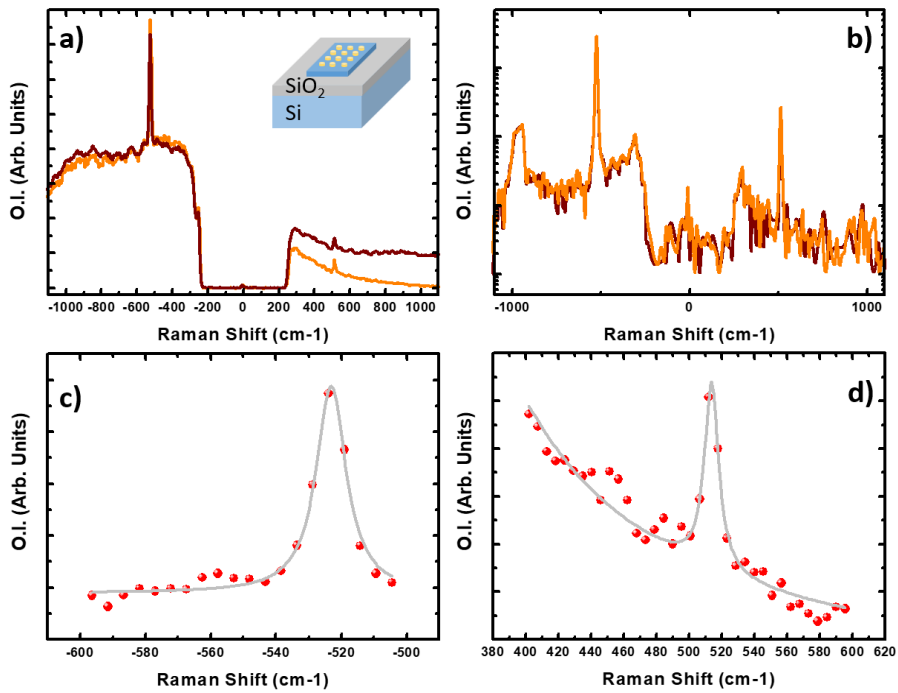

**Figure SI13**. (a) and (b) Raman spectra acquired on a patterned sample in arbitrary units and logarithmic arbitrary units respectively. (c) and (d) Lorentzian fitting to Stokes and antiStokes, respectively.

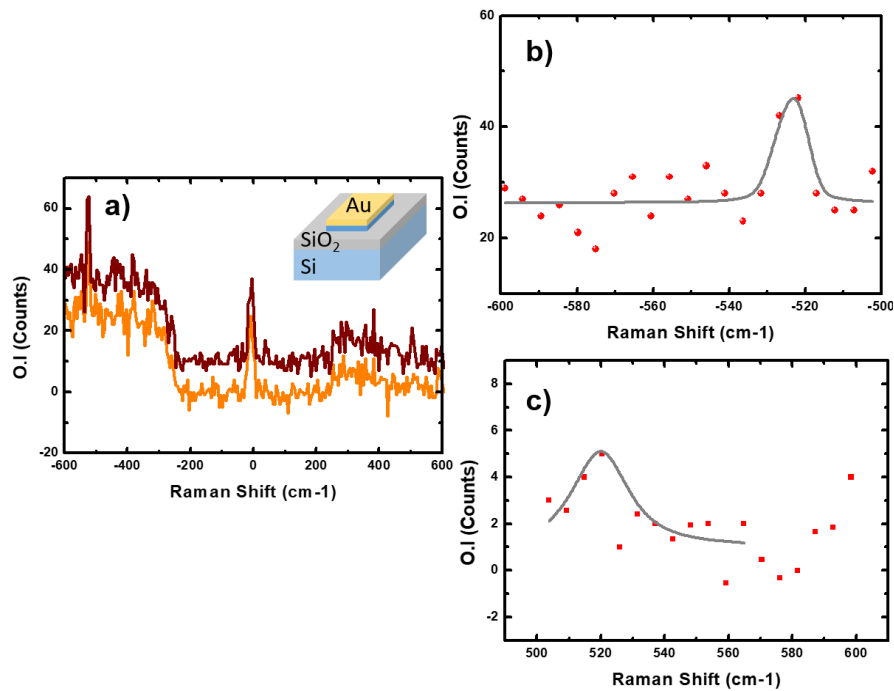

**Figure SI14** (a) Raman spectra acquired on a patterned sample in arbitrary units. (b) and (c) Lorentzian fitting to Stokes and antiStokes, respectively.

### Section SI7. Reflectivity measurements on the platform

In figure SI15a we show the reflection of a halogen lamp beam on a processed SoI substrate, at a distance of the silicon pads, where. Due to the thick silica layer and the high reflectivity of the silicon underneath the typical broadband spectra of the lamp is convoluted with multiple reflection peaks. This spectrum is used to normalize the reflectivity of the different kind of arrays fabricated in our thermoplasmonic platforms. As shown in Fig. SI15b the number of reflectivity features increases when measuring on the silicon pad. However, there is not a clear signature that we could attribute to a nanoantenna array. Because of this, the experimental demonstration of our proposal required of a strong numerical support and a bunch of reference samples.

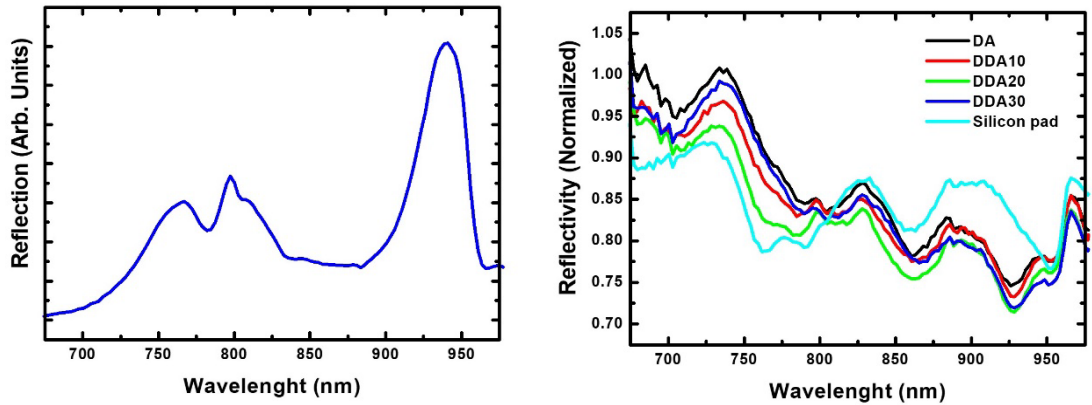

**Figure SI15** a) Reflection of the processed SoI. b) Reflectivity of different silicon pad with different kind of nanoantenna arrays.